\documentclass[aip,rsi,reprint,graphicx,twocolumn]{revtex4}
\usepackage{epsfig}
\usepackage{graphicx}% Include figure files
\usepackage{dcolumn}% Align table columns on decimal point
\usepackage{bm}% bold math
\begin{document}
%\preprint{APS/123-QED}
\title{ Interfering Coherent Lightsheets Assisted Structure Synthesis (iCLASS) Technique for Nanofabrication }% Force line breaks with \\

\author{ Kavya Mohan and Partha Pratim Mondal  }
 %\altaffiliation{}%Lines break automatically or can be forced with \\
 %\email{partha@fisica.unige.it}
\email[Corresponding author: partha@iap.iisc.ernet.in]{}
\affiliation{%
Nanobioimaging Laboratory, Department of Instrumentation and Applied Physics, Indian Institute of Science, Bangalore 560012, INDIA
}%
\date{\today}% It is always \today, today,
             %  but any date may be explicitly specified
%\doublespacing   
        
\begin{abstract}

We proposed and demonstrated a light-sheet based plane-selective fabrication technique that enables fabrication of nano-electronic/nano-fluidic components (nano-wires, nano-waveguides, nano-gratings and nano-channels) with specificity, selectivity on a user-defined patterning area. The technique is termed as, interfering coherent lightsheet assisted structure synthesis (iCLASS). iCLASS use specialized $2\pi$-geometry (consists of two opposing cylindrical lenses) that facilitates interference of counter-propagating light-sheets at a common geometrical focus. This gives rise to 1D interference pattern with a feature size of less than diffraction limit. A commercially available S1813 photoresist coated on a cleaned glass substrate is subsequently exposed to the light-sheets pattern (visible light). This is followed by the development of photoresist film to imprint the 1D nano-pattern. Experimental study and analysis shows an inter-feature spacing and feature-size of approximately, $\lambda/2$. Investigation show that the light-dose interaction-time ($\tau_{exp} +\tau_{dev}$) play crucial role in determining the feature-size and quality of fabricated nano-pattern array. The proposed iCLASS technique may find applications in nanoscience and nanotechnology. \\        
% ------------------------------------------------------------    

\end{abstract}

\maketitle

Next generation devices and machines demand onsite fabrication of nanoscale components. These complex systems will see an inter-mixing of nanoelectronics, nanophotonics and nanofluidics components for maneuvering complex engineering tasks. In particular, techniques that facilitate onsite-fabrication in remote locations (such as, outer space where reachability is difficult, costly and resources are limited) are of paramount importance for future development. \\

\begin{figure*}
\begin{center}
\includegraphics[scale=0.85]{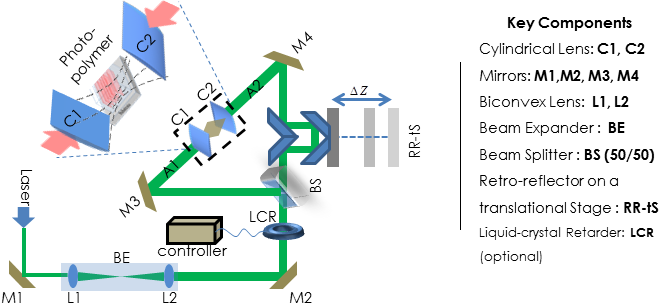}
%\rule{35em}{0.5pt}
\caption{Schematic diagram of the actual experimental setup for iCLASS system. }
\end{center}
\end{figure*}

\begin{figure}[htbp]
\begin{center}
\includegraphics[scale=0.5]{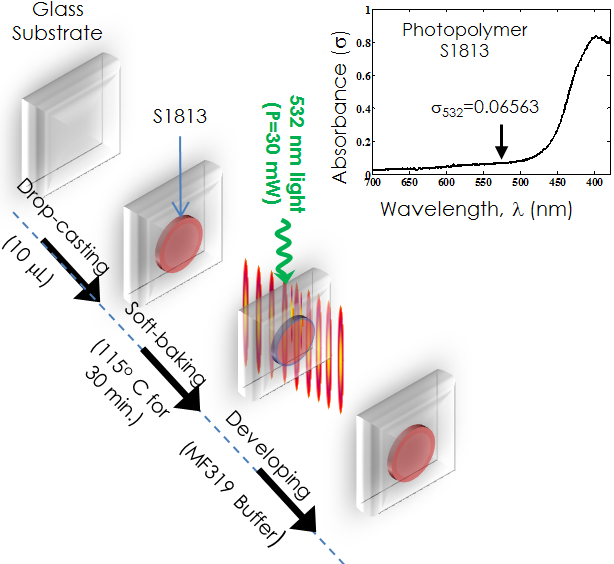}
%\rule{35em}{0.5pt}
\caption{Schematic diagram showing the intersection of light sheets with the photoresist film.  The UV-Vis absorption spectra of Michrochem S1813 is also shown. }
\end{center}
\end{figure}
% Figure ends*********************************** 

Optical lithography has gained interest over the past few decades. It has a key role in the success of semiconductor industry. Existing lithography techniques can be broadly categorized as, mask-based (photolithography and its variants) and mask-less (direct laser writing, laser interference lithography, STED lithography, focused ion beam lithography) [2,3,4,5,6,7]. These techniques are widely used even though they are complex, bulky and not miniaturizable. A major constraint with the existing light based nanolithography technique is that the feature size is limited by Abbe's diffraction limit because it uses light to transfer the pattern on the photosensitive specimen.  However, STED based techniques have show to achieve resolution beyond diffraction limit for nanolithography \cite{fischer2013}\cite{Klar}. Other key issue associated with nanolithography is the lack of control over the patterning area. Existing techniques such as laser interference lithography (LIL) is considered as a simple and cost-effective technique for the generation of periodic structures. In LIL, intensity of the interference pattern produce by the small number of coherent beams is recorded on the photoresist and is subsequently developed by the well known chemical protocols \cite{Dill}\cite{Lim2006}. In LIL the separation between two successive features is approximately, $\lambda/2n sin\theta$, $\theta$ being the inclination angle \cite{sreekanth}\cite{Lu2010}. Although LIL is limited to the generation of periodic structures, it has found wide range of applications in photonic crystals \cite{pelton}, magnetic storage devices \cite{farhoud} , biosensors \cite{Li} , diffraction gratings \cite{Xie} and broad band reflectors \cite{Seo}.\\

Light-sheet based technique has been proposed for fabricating simple micro/nano-structures \cite{kav2,kav3, mohan2016}. This technique is well-suited due to its specificity, plane-selectivity and maneuverable patterning area. The technique use coherently-illuminated counter-propagating interfering light-sheets for fabricating nanostructures. A distinct feature of this technique is the $2\pi$-geometry that allows single-shot fabrication of an entire specimen plane in a 3D volume. The schematic diagram of light-sheet based nanolithography system is shown in Fig.1. A coherent laser light source (Laserglow LSS-052, 532 nm,  coherence length, $l_{coh} >280~m$) is appropriately expanded using a beam-expander (BE) and directed towards the beam-splitter (BS) after passing through liquid crystal retarder (LCR). LCR may be used to alter the polarization of incident light and study its effects on the fabricated pattern. The coherent light is split into two parts: first part directly goes to the mirror M3 while the second part goes to the mirror M4 after passing through the retro-reflector (RR-tS) placed on a positioner. RR-tS enables necessary change in path-difference that is essential for interference. The beams are then reflected by mirrors, M3 and M4 to the cylindrical lens, C1 and C2 resulting in a pair of counter-propagating light-sheets. Note that path of the beam in the illumination arm A2 is altered using RR-tS to enable interference of light sheets at the common geometrical focus of C1 and C2 (see, Fig.1). The specimen to be patterned is placed at the focus and subsequently chemical protocols are followed to develop the pattern. \\

The scalar theory behind the proposed iCLASS system is essentially based on the interference of counter-propagating light waves. In general, the interference of two counter-propagating waves ($e^{ikx}$ and $e^{-ikx}$) gives rise to an intensity distribution at the common geometrical focus, $I=2I_0[1+\cos(\Delta\phi)]$, where, $\Delta\phi = (\phi_1 -\phi_2)=2kx$ is the phase-difference, $\Delta x =2x$ is the path-dfference and $k$ is the wave-number of light. The maximum (of the intensity distribution) occurs when, $\Delta\phi=2kx=2n\pi ~\Rightarrow x=n\lambda/2$, where, $n$ is an integer. So the period of the resulting pattern is, $\lambda/2$, further suggesting that a selective plane can be patterned with a period of $\lambda/2$. A much more precise and explicit expression for computing the field at the focus can be found using vectorial theory of light-sheet \cite{sub1}\cite{mohan2016}\cite{kav2}\cite{kav3}. In addition, $iCLASS$ offers user-defined patterning area (decided by the dimension of the interfering light-sheets assuming complete overlap) in a 3D volume with plane selectivity. \\

The goal is to exploit selective plane illumination property of interfering light-sheets in a $2\pi$-geometry for transferring the nano-pattern on a photopolymer system. The schematic diagram of the experimental setup is shown in Fig.1. Two counter-propagating wavefronts are made to interfere at the common geometrical focus of the cylindrical lenses, C1 and C2. A commercially available positive photopolymer (Michrochem-S1813) is exposed to the interfering pattern. Specific protocol for development of the photopolymer film is followed as shown in Fig.2. Corresponding photopolymer sensitivity is also shown for reference. UV-Vis absorption spectra is recorded using a UV-Vis spectrometer ( SHIMADZU, UV- 2600). The absorption spectrum was recorded by diluting the photopolymer mixture in MF319 ($1 \mu l$ in 1ml of MF319) and by keeping MF319 as a reference. S1813 has relatively weak absorbance for visible region of the electromagnetic spectrum. This ensures depth-penetration over the thickness of photoresist film. We have used relatively higher laser power to account for the decreased photosensitivity. Experiments were performed with a coverglass ($1cm \times 1cm$) immersed in Chromic acid cleaning solution  at $70^o \pm 2^o C$ for 15 min to remove organic and inorganic contaminants. Subsequently, the substrates were coated with Michrochem-S1813 photoresist by drop-casting method. Samples were soft-baked at $115^o \pm 2^o C$ for 30 min and developed using MF319 buffer (see, Fig.2). $10 \mu l$ of S1813 on a $1cm \times 1cm$ coverglass has resulted in a film thickness of  $ \approx 300 \mu m$. Thickness measurements were carried out on a dektak surface profiler. S1813 is tuned for g-line exposure(436nm). Hence determining a optimum exposure dose for 532nm illumination is necessary for achieving patterns on the photoresist. Note that,  the exposure dose varies with the thickness of the photoresist film. \\

%Computational studies were carried out to determine the field at the geometrical focus using vectorial theory of light \cite{sub1}\cite{kav3}. Studies show the formation of 1D nano-patterns (such as, nano-wire/nano-grating/nano-channels) of dimension 280 nm (see, Fig.3).  \\

%\begin{figure*}[htbp]
%\begin{center}
%\includegraphics[scale=0.5]{Fig3-Nat-Comm.png}
%%\rule{35em}{0.5pt}
%\caption{Cartoon representing the steps involved in the photoresist patterning. Step-3 shows the illumination of photoresist with the interference pattern. Half of the length of light sheets is made to intersect with the photoresist.Exposure dose is determined by keeping the illumination power and developing time constant. wavelength of exposure ($\lambda_{exp}: 532nm$), $T_{exp}$: exposure time, $T_{dev}$: developing time.Images are taken in a Nikon eclipse LV100 with an objective of 0.4 NA and 20 X magnification.}
%\label{fig:steps}
%\end{center}
%\end{figure*} 

\begin{figure}[htbp]
\begin{center}
\includegraphics[scale=0.45]{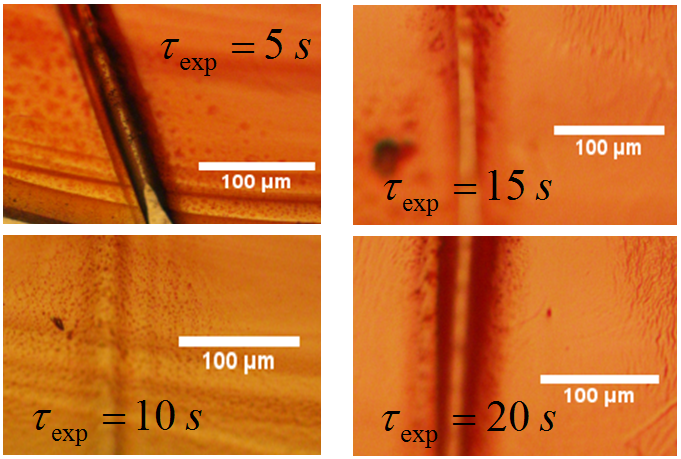}
%\rule{35em}{0.5pt}
\caption{Optical microscopy images taken using a Nikon eclipse LV100 with an objective of 0.4 NA and 20X magnification.}
\end{center}
\end{figure} 

\begin{figure}[htbp]
\begin{center}
\includegraphics[scale=0.85]{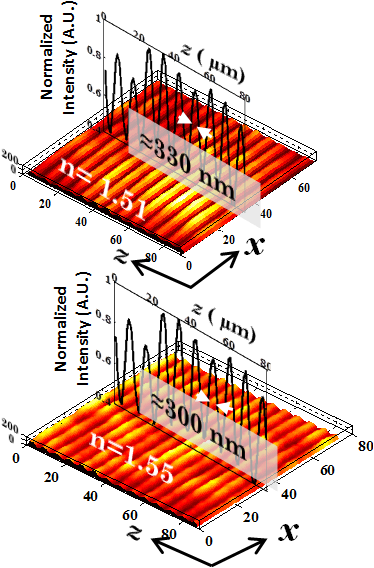}
%\rule{35em}{0.5pt}
\caption{High resolution optical images of 1D nano-patterns in  Glycerol (RI: 1.51) and Agarose-gel matrix (RI: 1.55).}
\label{fig:AFM}
\end{center}
\end{figure} 

\begin{figure*}[htbp]
\begin{center}
\includegraphics[scale=0.3]{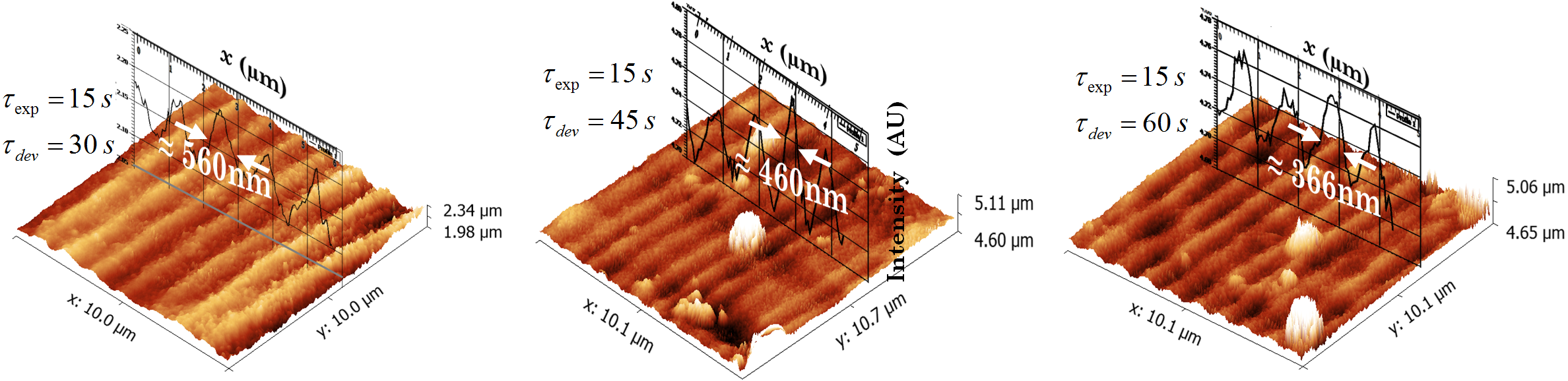}
%\rule{35em}{0.5pt}
\caption{3D  AFM images obtained for variable developing time ($\tau_{dev}$) of 30s, 45s and 60s and exposure time($\tau_{exp}$) of 15s. Intensity line plots are taken across the pattern showing variable channel-width. Illumination power used is 18 mW.}
\label{fig:AFM}
\end{center}
\end{figure*} 

In order to determine the optimum light-dose interaction-time, we have carried out experiments for different values of exposure time by keeping the illumination intensity and the developing time constant. Experiments are carried out for a light of wavelength, $\lambda=532nm$ at an illumination power of 18 mW. The samples were mounted perpendicular to the direction of illumination as shown in Fig.2. The exposure time is varied from 5s to 20s in steps of 5s. Exposed samples were developed for 1 min in MF319 buffer solution. Developing mainly involves the diffusion of the solvated polymer chains into the developing solution. Developing time generally depends on the thickness of the photoresist film, exposure dose, temperature of the developing solution and also on the geometry of the patterns. The substrates were subsequently rinsed in ample distilled water for 10 minutes and left to dry in air. All the procedures were performed at a room temperature of $18^o \pm 2^o C$ under low light conditions. The results obtained are as shown in Fig.3. It is evident that an exposure time of $\tau_{exp}=15 ~secs$ is optimal for fabrication. This suggests that optimal exposure time is essential for quality with an effective balance of exposure time and development time. Fig.4 show high resolution optical images (obtained using orthogonal detection system equipped with Olympus 100X 1.3NA objective lens) of the developed pattern that reveals 1D nano-channels of sizes approximately, 330 nm and 300 nm for two different refractive indexes, 1.51 (Glycerol) and 1.55 (Agarose Gel) respectively (similar to Ref.\cite{mohan2016}). \\

Fig.5 shows 3D AFM images of the photosensitive specimen exposed for 15 secs (exposure time) and with varying developing time. All these images are recorded at the periphery of the photoresist in a non-contact mode. Thickness of the photoresist is not uniform at the periphery. Hence it becomes easier to record the interference patterns on a substrate that is kept perpendicular to the counter-propagating light-sheets. AFM images clearly shows that the contrast is low for $\tau_{dev}=30~secs$ when compared to images for 45 secs and 60 secs. This is purely due to insufficient developing time. The intensity profile plots taken across the 1D pattern is also shown. For the profile plots we have calculated the average peak width for each case. The average peak-width for light-dose interaction time ($\tau_{exp} +\tau_{dev}= 15s+60s =75s$) is found to be approximately, 366 nm which is in a close agreement with the theoretical predictions, $\lambda/2 = 266 ~nm$. We found that the increase in light-dose time results in other undesirable effects. We believe that other factors related to the chemical properties of photoresist photopolymer play active role in determining the effective width of the fabricated 1D nano-patterns. \\
   
In conclusion, light-sheet based iCLASS fabrication technique is an alternate route to fabricate 1D nano-patterns. The key optical elements are, coherent light-source, photo-sensitive photopolymer and opposing cylindrical lens arranged in a $2\pi$-geometrical optical setup for light-photopolymer interaction. iCLASS system is highly miniaturizable and cost-effective. Selective plane illumination raises the hope for on-site manufacturing of nano-circuitry for integrated 3D chips, nanofluidic channels and nano-optoelectronic components. Advances in onsite nano-fabrication techniques are expected to fuel rapid progress in the manufacturing of complex miniaturized systems especially in challenging locations (intersteller space and rural/isolated remote areas). \\

{\bf{Acknowledgment:}}
Both the authors contribute equally to this work.

\end{document}